# Trying to bridge the gap between skyline and top-k queries

Alessandro, AP, Pindozzi
Politecnico di Milano
Milan, Italy
alessandro.pindozzi@mail.polimi.it

**Abstract**

There are two most common paradigms that are used in order to identify records of preference in a multi-objective settings, one relies on dominance, like the skyline operator, the other instead, on a utility function defined over the records' attributes, typically using top-k queries.

Although they are very popular, we have to take in account their main disadvantages, which bring us to describe three hard requirements: personalization, controllable output size, and flexibility in preference specification. In fact Skyline queries are simple to specify but they are not equipped with any means to accommodate user preferences or to control the cardinality of the result set. Ranking queries adopt, instead, a specific scoring function to rank tuples, and can easily control the output size, but it is difficult to specify correctly the weights of this scoring function in order to give different importance to the attributes.

In this paper we describe three different approaches which try to satisfy the three hard requirements mentioned above embracing the advantages either of the Skyline queries or of the ranking queries. These approaches are namely: Flexible Skyline, ORD-ORU and UTK.

**Keywords:** Top-k query, Skyline, ORD-ORU, UTK

## 1 Introduction

In the era of ubiquitous access to the Internet, users are presented with numerous alternatives to cover their everyday needs. Choosing from the available alternatives generally entails the consideration of multiple, often conflicting aspects. All these aspects have to be simultaneously met in the best possible way according to the user's preferences. This is the so-called multi-objective optimization [9].

As we described in the Abstract, for a large set of alternatives (i.e., d-dimensional records) there are two main paradigms to determine those of most interest to the user, which are dominance and ranking by utility, but here we give more precise definitions of them.

Regarding the first paradigm, we can say that a point (or a record, as in our case) dominates another point (another record) if it is as good or better in all dimensions and better in at least one dimension. The skyline is defined as those points (records) which are not dominated by any other point [1], while the k-skyband is the set of points (records) which are dominated by at most k-1 points (records) [14]. Given these definitions, we are going to describe the main shortcomings of the dominance paradigm and briefly why they are so important [7]:

1. it is not personalizable, reporting the same result for every user.
   Personalization (i.e., serving the specific preferences of an individual user) is a hard requirement for decision support;

2. its output size (i.e., the number of reported records) is uncontrollable.

The time it takes for a person to make a decision as a result of the possible choices: increasing the number of choices will increase the decision time logarithmically [10, 8]. Hence, another hard requirement is for output-size specified (OSS) operators.

The second paradigm, instead, ranking by utility, in the form of top-k query, is both personalized and OSS. However a ranking query's result is heavily influenced by the choice of weights within a scoring function. This is a difficult task, since it is hard to predict the effects of changing one or more parameters on ranking. The attributes weights can be be either input directly by the user or somehow mined (e.g. via machine learning) [6], but in the first case we have to consider that a user is not able to give the right weights precisely, while in the second one we must take into account that the mined weights are only estimates.

Table 1 summarizes pros and cons of multi-objective optimization approaches Skyline and Ranking queries, in light of what is written above.

| Evaluation Criteria | Ranking | Skyline |
| --- | --- | --- |
| Simplicity of formulation | No, it is difficult to define or mine the attributes weights of the scoring function. | Yes, we don't need to weight any attribute. |
| Control of result cardinality | Yes, top-k queries return exactly k elements. | No, we cannot control the cardinality of the set of nondominated records. |
| Overall view of interesting results | No, providing only k results may result in missing some interesting results. | Yes, the set of non-dominated records are considered interesting results. |
| Relative importance of attributes | Yes, giving different weights to the attributes in the scoring function. | No, the skyline does not provide any way to give more importance to some attributes than the others. |

Table 1: Pros and cons of multi-objective optimization approaches

In this paper we describe and compare three different approaches which are aimed at bringing together the strong points of both paradigms (dominance-based and ranking by utility), while avoiding their drawbacks, which are: Flexible Skyline, ORD-ORU and UTK.

## 2 Flexible Skyline

Before starting explaining what is flexible skyline, let's assume we have a relational schema $R^+(A_1,...,A_d)$ with $d \leq 1$. Then we assume that the domain of each attribute $A_i$ is [0,1], since each numeric domain could be normalized in this interval. We consider higher values to be better than lower ones, but the opposite convention would of course also be possible. A tuple $t$ over $R$ is a function that associates a value $v_i$ in [0,1] with each attribute $A_i$; $t$ is also written as $<v_1,...,v_d>$, and each $v_i$ may be denoted by $t[A_i]$. Given the geometric interpretation of a tuple in this context, in the following, we sometimes also call it a point. An instance over $R$ is a set of tuples over R. In the following, we refer to an instance $r$ over $R$[5, 4].

### 2.1 Skyline

At first, let's recap the concept of dominance and Skyline, which is important in order to understand *flexibleskyline* and allows to make explanatory comparisons.

Given two tuples in R $t$ and $s$, $t$ *dominates* $s$ $t \prec s$, if:

1. $\forall i. \geq i \geq d \rightarrow t[A_i] \geq s[A_i]$

2. $\exists j. 1 \geq j \geq d \wedge t[A_j] < s[A_j]$

The *skyline* of $r$, denoted by $SKY(r)$, is defined as

$$SKY(r) = \{t \in r | \nexists s \in r.s \succ t\} \tag{1}$$

An equivalent definition of skyline may be derived by resorting to the notion of monotone scoring functions, i.e., those monotone functions that can be applied to tuples over R to obtain a non-negative value representing a score[5, 4]. Therefore, as formally shown in Reference [3], the skyline of $r$ can be equivalently specified as:

$$SKY(r) = \{t \in r | \exists f \in MF. \forall s \in r.s /= t \rightarrow f(t) > f(s)\} \tag{2}$$

where *MF* is the set of all monotone scoring functions.

## 2.2 Flexible Skyline operators

In light of what is written above, we can now describe *flexible skyline*, distinguishing two *flexible skyline* operators: *non-dominated flexible skyline* and *potentially optimal flexible skyline*. The behavior of these two operators is the same as SKY, but applied to a limited set of monotone scoring functions $F \subseteq MF$ [5]. We also assume that $F$ is not empty and finite.

We now extend the notion of dominance introduced in the Formula 1 to take into account the set of scoring functions we have just described. Given two tuples in R $t$ and $s$, $t$ F-dominates $s$:

$$t \succ_F s, if \forall f \in F. f(t) \geq f(s). \tag{3}$$

There are several properties about the *F-dominance*, which regard preservation or transitivity, but it is not the purpose of this paper to describe them all in detail. However, the reader can find such deeper notions in Reference [5].

Now we give formal definitions about *non-dominated flexible skyline* and *potentially optimal flexible skyline*. In all these we consider $F \subseteq MF$ a set of monotone scoring functions.

With Definition 3 at hand, we are now ready to introduce the first *flexible skyline operator*, called *nondominated flexible skyline*, which consists of the set of *non-F-dominated* tuples in $r$, as specified in the Definition 4 below [5, 4]:

$$ND(r;F) = \{t \in r | \nexists s \in r.s \succ_F t\}. \tag{4}$$

It is easy to notice that the Definition 1 of SKY(r) and 4 coincide except for the symbol $\succ$, which is replaced by $\succ_F$.

The second flexible skyline operator, called potentially optimal flexible skyline, returns the tuples in $r$ that are best (i.e., top-1) according to some scoring function in F, as specified in Definition 5 below [5, 4]:

$$PO(r;F) = \{t \in r | \exists f \in F. \forall s \in r.s \neq t \rightarrow f(t) > f(s)\}. \tag{5}$$

Also in this case it is easy to notice that the Definition 2 of SKY(r) and 5 coincide except for *MF*, which has been replaced by *F*.

We now describe the main properties of *ND* and *PO*.

As we said before $F \subseteq MF$ (the set of all monotone scoring function), so it is trivial that [5, 4]:

$$PO(r;MF) = ND(r;MF) = SKY(r). \tag{6}$$

While, considering the general case, the following relationship hold [5, 4]:

$$PO(r;F) \subseteq ND(r;F) \subseteq SKY(r). \tag{7}$$

We define *F-dominance* region of a tuple *t* under a set of monotone scoring functions *F*, *DR(t;F)*, as the set of all points $[0,1]^d$ that are *F*-dominated by *t* [5, 4]:

$$DR(t;F) = \{s \in [0,1]^d | t \succ_F s\}. \tag{8}$$

More properties and definitions can be found in References [5] and [4].

We report in Table 2 below the Definitions 4 and 5 of the two *flexible-skyline operators* and of the classic *skyline* in order to better highlight the differences between them:

| Skyline with dominance $\succ$ | $SKY(r) = \{t \in r | \nexists s \in r.s \succ t\}$ | Skyline with scoring function | $SKY(r) = \{t \in r | \exists f \in MF. \forall s \in r.s/= t \rightarrow f(t) > f(s)\}$ |
|---|---|---|---|
| Non-dominated flexible skyline | $ND(r;F) = \{t \in r | \nexists s \in r.s \succ_F t\}$ | Potentially optimal flexible skyline | $PO(r;F) = \{t \in r | \exists f \in F. \forall s \in r.s/= t \rightarrow f(t) > f(s)\}$ |

Table 2: Definitions of skyline and f-skyline operators

# 3 UTK

The traditional top-k query receives as input a dataset with *d*-dimensional records, and a vector *w* of *d* weights that specify the relative significance of each dimension (data attribute) for the user. The score of a record is defined as the weighted sum of its attribute values. The k highest-scoring records form the output of the top-k query. The weight vector *w* represents the user's preferences. We have already described how much can be difficult to find the right balance of these weights.

Another way to deal with this issue consists of expanding the weight vector into a region, and report the additional options identified by *UTK*. UTK needs three inputs[13]:

1. a dataset *D*, a positive integer *k*, and a region *R*. Each record $p \in D$ includes *d* values, i.e., $p = (x_1, x_2, ..., x_d)$;

2. a *weight vector* $w = (w_1, w_2, ..., w_d)$. We assume that $w_i \in (0,1)$ for each $i \in [1,d]$ and that $\sum_{i=1}^{d} w_i = 1$.

   This last condition allows to drop one weight, in fact we can derive $w_d = \sum_{i=1}^{d-1} w_i$, and thus reduce the domain of *w* to a (*d* - 1)-dimensional space, called the *preference domain*;

3. a region *R* in preference domain. For ease of presentation, we assume that it is an axis-parallel hyperrectangle.

Moreover, the score of each record *p* is derived, with the weighted vector *w* as $S(p) = \sum_{i=1}^{d} w_i \cdot x_i$. Also in this case, as we did for *f-skyline* (2), we assume higher values to be better than lower ones.

Mouratidis and Bo Tang in their article [13] distinguish distinguish two UTK versions. $UTK_1$ reports the set of exactly those records that may rank among the top-k when the weight vector lies inside *R*. "Exactly" here means that the reported set is minimal, i.e., for every record *p* in it, there is at least one weight vector in *R* for which *p* belongs to the top-*k* set. The second version, $UTK_2$, reports the exact top-*k* set for every possible weight vector in *R*. While there are infinite possible vectors in *R*, the output is a partitioning of *R*, where each partition is associated with the exact top-*k* set when *w* lies anywhere inside that partition.

Mouratidis and Bo Tang define the concept of *r-dominance* as follows: "Given a region *R* in the *preference domain*, we say that record *p* r-dominates another record *p'* when $S(p) \geq S(p')$ for any weight vector in *R*, and there is at least a weight vector *R* for which $S(p) > S(p')$."

The r-dominance concept is fundamental, in fact any record that is *r*-dominated by *k* or more other records cannot be in the *UTK*$_1$ result [13].

## 4 ORD-ORU

We consider that the available options are represented as *d*-dimensional records *r* =< $x_1, x_2, ..x_d$ > in a dataset D. We make the convention that the larger the attributes the better.

Given a preference vector *v* of non-negative weights $w_i$, the utility score of a record *r* is defined as their inner product, i.e., $U_v(r) = \sum_{i=1}^{d} w_i \cdot x_i$. Accordingly, the top-*k* result comprises the *k* records with the highest scores. We make the same assumption as for UTK: $\sum_{i=1}^{d} w_i = 1$, and so we assume preference vectors, called *preference domain*, is the unit (d-1)-simplex in a space whose *d* axes correspond to the $w_i$ values [12]. In other words, for *d* = 3, the *preference domain* is an equilateral triangle, and any valid preference vector is represented as a vertex in that triangle. For *d* = 4, the *preference domain* is a tetrahedron, and so on.

Let *w* be a best-effort estimate of the user's preference vector, the so-called *seed*, and consider the preference vectors *v* within distance *d* from *w*, i.e., where |*v* − *w*| ≤ *d*. If a record $r_i$ scores at least as high as another $r_j$ for every such vector *v*, and strictly higher for at least one of them, we say that $r_i$ *ρ*-dominates $r_j$ [12]. All the records that are *ρ*-dominated by fewer than *k* others form the *ρ*-skyband.

In this context we can define the two operators ORD and ORU which purpose is to have both to have a high personalization of the result set based on the user's preferences and have a specified output size.

ORD is an acronym, which highlight its OSS property, relaxed input, and stronger dominance-oriented flavor. It is defined by Kyriakos Mouratidis, Keming Li and Bo Tang [12], given the seed vector *v* and the required output size *m*, as the records that are *ρ*-dominated by fewer than *k* others, for the minimum *ρ* that produces exactly *m* records in the output.

Since a record can not belong to the top-*k* result for any preference vector if it is not in the *k*-skyband [1], it follows that the *ρ*-skyband is a subset of the *k*-skyband. Moreover, it can happen that a record does not dominate another one in the traditional sense, but it scores higher for *w*.

In this context we can introduce the concept of *inflection radius*: considering a tuple $r_i$, its inflection radius is the value of *ρ* for which $r_i$ is *ρ* dominated by less than k other tuples. Hence, computing the entire *k*-skyband, and deriving for each tuple, its inflection radius, we can output the *m* tuples with the smallest radius.

ORU, instead, as ORD aims to point out its OSS property, relaxed input, but the last letter specifies how it follows more closely the ranking by utility paradigm. In the same reference cited above [12], it is stated that, given the seed vector *v*, ORU reports the records that belong to the top-*k* result for at least one preference vector within distance *ρ* from *w*, for the minimum *ρ* that produces exactly *m* records in the output.

In order to better explain ORU we have to define the so-called *convex-hull*. The *convex-hull* of *D* is the smallest polytope that encloses all its records. It comprises facets, each defined by *d* extreme vertices (records) in general position. The *norm* of a facet on the hull is the normal vector to that facet whose sum of coordinates is 1, and is directed towards the exterior of the hull. The top record for a preference vector *v* is the one met first by a hyper-plane normal to *v* that sweeps the data space from the top corner to the origin [12]. Thus, we call the *upper hull* or layer, the set of all facets of that *convex hull* whose normal vector is directed to the positive part of the plane, and last, we call *top-region* of *t*, the part of the domain in which *t* is the tuple that has the highest score. We can give another definition before talking about the computation of ORU. If we consider a record *r*, in a certain layer, we say that the set of the records adjacent to *r* are those records of the same layer which are vertices of the facets (of that layer) of which *r* is the other vertices. Regarding the computation of ORU, if we assume that a certain tuple *t* is the top-*i* result for a preference vector *v*, the following best result (top-*i*+1) will be among the adjacent tuples in the same layer or in the tuples in the next layer whose top-region overlap the one of *t*.

# 5  A trivial example

In this section we provide a simple example which aim is to help the reader to better understand the behaviour of the *f-skyline*, UTK and ORD-ORU. We preferred to analyze a unique example rather than a different one for each approaches, in order to point out the main differences between them. Moreover we can see also the results given by top-*k* and classic skyline, compared with the ones of *f-skyline*, UTK and ORD-ORU.

Let's assume that Bob decides to buy a new computer, and that he takes in account two parameters: the number of cores and the RAM capacity. Obviously his preference is to buy the best computer in terms of number of cores and the RAM capacity. In other words the computer which has the maximum value of both these attributes.

In Table 3 below the list of the available computers.

| ComputerID | Core | RAM(Gb) |
|---|---|---|
| p1 | 2 | 6 |
| p2 | 7 | 5 |
| p3 | 8 | 3.2 |
| p4 | 10 | 2.9 |
| p5 | 12 | 0.8 |
| p6 | 2 | 4 |
| p7 | 5 | 2.7 |
| p8 | 9 | 1.2 |
| p9 | 1 | 2 |
| p10 | 3 | 0.5 |

Table 3: Available computers, with their RAM capacity and number of cores

The available options are represented as *d*-dimensional records $p = <p_1, p_2, .. p_d>$ in a dataset D where $d \in [1, 10]$.

## 5.1  Top-k

Using the Top-*k* criterion, we can define a scoring function as follows: $S(p) = w_c \cdot Core + w_r \cdot RAM$. We set *k=1*, so that the top-*k* query will return the best possible computer, in fact he top-*k* set includes the *k* records with the largest scores [11].

At first, we can assume that Bob cares more the number of cores than the RAM capacity, for example $w_c = 0.7$ and $w_r = 0.3$. As you can see in Figure 1, the best computer, in this case, is p5.

Instead, let $w_c = 0.2$ and $w_r = 0.8$. You can notice that the best solution in this case is p2 (see Figure 1).

It comes out that top-*k* is closely related to weights and to their variations. That's why, as previously explained, giving right weights to the attributes is of fundamental importance, but it is an arduous task.

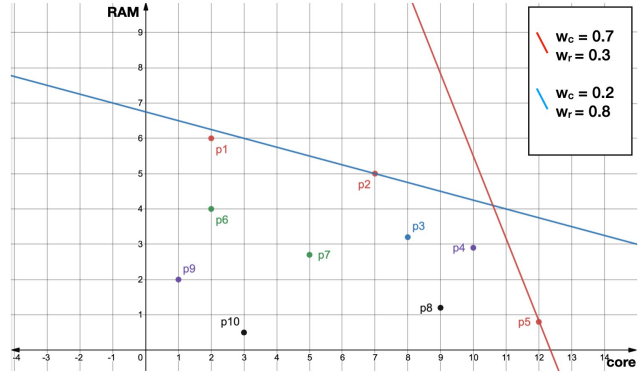

Figure 1: Top-*k* queries

## 5.2 Skyline

Now we try to use classic skyline queries. As we know, the records which form the *skyband* are the ones which are not dominated by other records. It is obvious that the points that are not in the skyline are not best options, regardless of the criterion chosen (if Bob give more importance to the cores or to the RAM).

As you can see in Figure 2 the points which belong to the skyline are p1, p2, p3, p4, p5. Also in this case the limitations of this approach emerge: it is not personalizable, in fact no weight has been assigned to the attributes, and its output size is not controllable, since it is not possible to predict how many records will be part of the skyline.

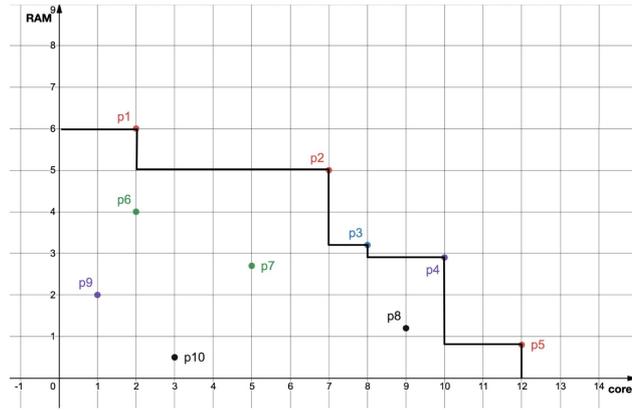

Figure 2: Skyline queries

## 5.3 Flexible Skyline

In the previous subsection (5.2) we established that SKY(p)={p1, p2, p3, p4, p5}.

Now, let $F$ be the set of all the linear scoring functions of the form $f(x,y) = w_c \cdot Core + w_r \cdot RAM$ such that $w_c \geq w_r$.

We note that p2 $f$-dominates p1 ($p2 \succ_F p1$), in fact from $f(p2) \geq f(p1)$, which is $w_c \cdot 7 + w_r \cdot 5 \geq w_c \cdot 2 + w_r \cdot 6$, we obtain $w_c \cdot (7-2) \geq w_r \cdot (6-5)$, and because we are assuming that $w_c \geq w_r$, it is always satisfied. But p2 is $f$-dominated by p4 ($p4 \succ_F p2$), since $w_c \cdot (10-7) \geq w_r \cdot (5-2.9)$ is always satisfied for all $w_c$ and $w_r$. The same is true for p3, which is $f$-dominated by p4 ($p4 \succ_F p3$): we have that $w_c \cdot (10-8) \geq w_r \cdot (3.2-2.9)$ is always satisfied. It follows that $p1, p2, p3 \in/ ND(p;F)$. Instead, neither p4 nor p5 are $f$-dominated by other tuples in p, so they belongs to $ND(p;F)$. You can see the results in Figure 3(a).

We can also determine which are the potentially optimal records. As you can see in Figure 3(b), both p4 and p5 are potentially optimal. In fact, if we draw the lines defined by the scoring function in $F$, we notice that there can be two different results:

- if $w_c = w_r$, which means that Bob gives the same importance to number of cores and RAM capacity, the optimal result is p4, since it is the first point that meets the line;

- if $w_c \simeq 0.51$ and $w_r \simeq 0.49$, the optimal result are both p4 and p5;

- as $w_c$ grows (so for $w_c > w_r$, with $w_c > 0.51$), which means that Bob prefers a high number of cores than a high RAM capacity, the potentially optimal tuple is p5.

It emerges that, in this example, $PO(p : F) = ND(p;F) \subseteq SKY(p)$.

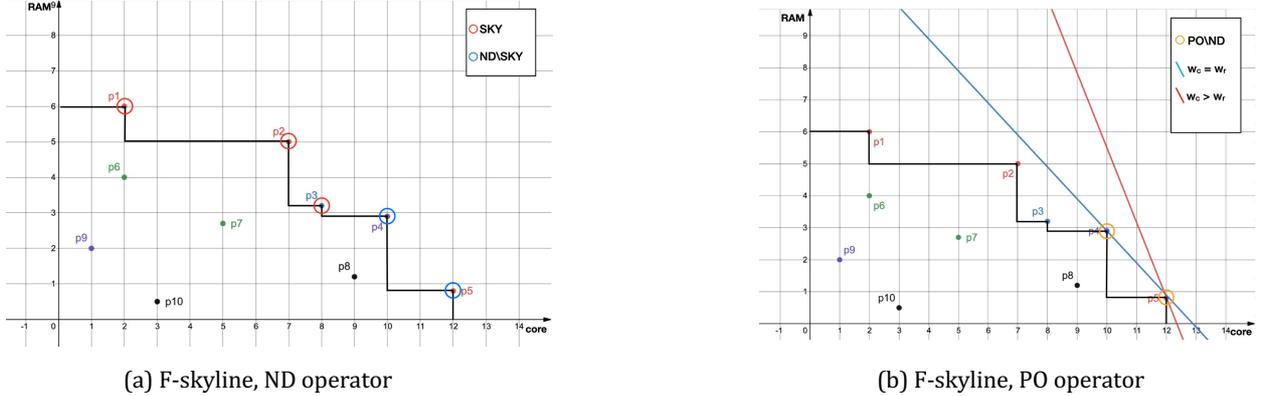

(a) F-skyline, ND operator  (b) F-skyline, PO operator

Figure 3: F-skyline operators

## 5.4 ORD-ORU

At first let's see how ORD works, but we consider only p1, p2, and p3 for simplicity of calculations and understanding. Let the seed vector be $w = [0.5, 0.5]$. The score of tuples p1, p2 and p3 will be respectively $U_w(p1) = 4$, $U_w(p2) = 6$ and $U_w(p3) = 5.6$.

It is clear that p2 and p3 score higher than p1, but they do not dominate it in the classical definition of dominance, since p1 has a better value of RAM (see Table 3).

Now we want to compute the inflection radius of p1 for k=2. Given the vectors $v_i$ of distance 1 from $w$, [0.5,1.5] [1.5,0.5], we calculate again the score of the three tuples that we are considering (scores are shown in Table 4). Using the vector [0.5,1.5], $U_{v1}(p1) = 10$, while $U_{v1}(p3) = 8.8$, so p3 score lower than p1. This is sufficient to say that p1 is dominated by less than 2 tuples for a radius of 1 which is its inflection radius.

| ComputerID | Score with the seed vector w | Score with the preference vector v=[0.5,1.5] |
|---|---|---|
| p1 | 4 | 10 |
| p2 | 6 | 11 |
| p3 | 5.6 | 8.8 |

Table 4: Scores of p1, p2, p3 with the seed vector and the preference vector

Now we can explain better how ORU works. In Figure 4, all the tuples of $D$ plotted and the the upper hull is drawn. The facets of that hull are the segments p1-p2, p2-p4, p4-p5, while v1, v2 and v3 are the corresponding normal vector. The record that scores the most for a given vector $w = (0.5, 0.5)$ is p4, so it is the top-1 record. If we consider an output size $m=2$, the first element will be p4 as we stated before, while the next record that will score the maximum for a given vector $v$ in $\rho$ distance from $w$ will be among the set of adjacent records, hence

{p2,p5} (the definitions of adjacent record has been given in 4), or among the set of the tuples in the next layer whose top-region overlap the one of p4, which is {p3, p8}.

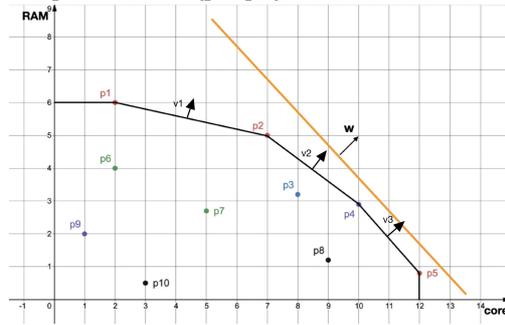

Figure 4: ORU - convex hull

## 5.5 UTK

Let's suppose we have a weight-vector $w = (w_c, w_r)$, that $w_c + w_r = 1$ and that $w_c \geq w_r$. The scoring function described in Section 3 can be written as: $S(p) = w_c \cdot Core + w_r \cdot RAM$. Since the weight-vector is 2-dimensional ($w_c$ and $w_r$), as explained in Section 3, the *preference domain* is 1-dimensional. Moreover, from $w_c + w_r = 1$ we derive $w_r = 1 - w_c$.

If we plot the scores of records $p \in D$ as functions of $w_c$, they are each mapped into a line (see Figure 5). We did this for five records, which are the ones corresponding to the output of the classic skyline (see 5.2). If we consider a certain value of $w_c$, and we draw a vertical line from this value, it meets the lines of the records in ascending order of $S(p)$. I.e., the top-$k$ set for $w$ comprises those records that correspond to the $k$ lines that are met last by the ray. If we consider the region in which can change $w_c$ (considering $w_c \geq w_r$), we have the following result:

- if $w_c = w_r$, which means that Bob gives the same importance to the number of cores and RAM capacity, the optimal result is p4, since its line is the first which meet the ray of $w_c$;

- if $w_c \simeq 0.51$, the ray meets for last simultaneously both the line of p4 and p5;

- if $w_c > w_r$ (if $w_c > 0.51$), which means that Bob prefers a high number of cores than a high RAM capacity, the line that first meet the vertical ray is the one of p5.

So UTK result for $k = 1$, so if we want only the best options, are the tuples that correspond to the part of the lines which are in the preference region (delimited by bold orange lines in Figure 5), which are p4 and p5.

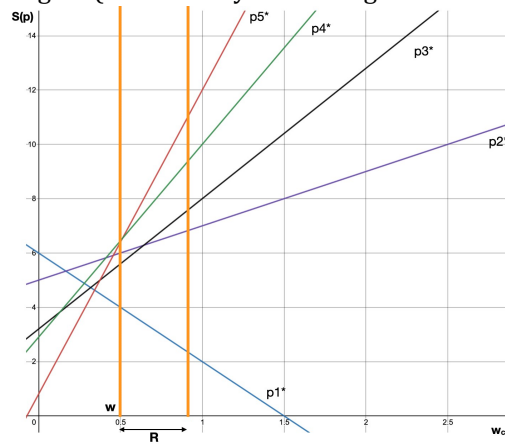

Figure 5: UTK result

# 6  A final review on the main aspects

In light of the example done in the previous Section we can compare the different approaches in terms of personalization, OSS and flexibility in preference.

- **f-skyline:** Ciaccia and Martinenghi introduced the notion of flexible skyline queries (F-skylines), aiming to extend the skyline framework with user preferences expressed by means of constraints on the weights of scoring functions [5, 4].In other words, the monotone scoring functions they describe are defined through constraints in order to achieve personalization.

    This approach does not allow to control directly the cardinality of the result, but narrows the field compared to classic skyline. Just look at the example (Section 5) to see that in the case of skyline the result was the set of tuples {p1, p2, p3, p4, p5}, while for $f$-skyline only p4 and p5 belongs to ND and PO. In our case only the 40% of the results of the classic skyline correspond to the output of $f$-skyline, which is a remarkable reduction.

    However, experiments done (Reference [5]) show how the cardinality of the results obtained with $f$ skyline can be even lower than those of the classic skyline queries: in case of ND the number of results correspond to the 9.8% of the amount of results obtained through skyline queries, while for PO are only the 1%.

    Moreover, in Reference [5] is suggested to use one of the methods developed for controlling the cardinality of the skyline orthogonally to $f$-skyline, where the result of ND and PO are too large.

- **UTK:** At first we want to highlight the relationship between UTK, skyband and convex-hull. We know that the $k$-skyband contains those records that are dominated by fewer than $k$ others, so it is a superset of the set of record that can be returned as a result of a top-$k$ query. We already stated that the *convex-hull* is the smallest convex polytope that encloses all records in a dataset (see Section 4).
    The top-1 record for any weight vector $w$ is guaranteed to be on the convex hull. This is used in the *onion* technique, which relates the layers of the onion to the convex-hull, to answer top-$k$ queries by looking only through $k$ layers (more information about this can be found in Reference [2]). Both the $k$-skyband and *onion* are related to UTK. They retain some records that cannot be in any top-$k$ set for any $w$. In UTK, where w is further bounded by R, the $k$-skyband and *onion* layers become even looser supersets of the records that could appear in the top-$k$ set. So they are used in algorithms for UTK as a filtering step, in order to determine for each retained candidate p whether it is part of the UTK result [13].

    F-skyline uses a family of scoring functions, which can be considered a weighted linear combination of the attributes of the records in a dataset, so it reports all records that could rank as the top-1 for any permissible weight combination. In other words, flexible skyline is strictly related to $k$=1.

    UTK also considers a scoring function which must be monotone to the attributes of the records and linear to the weights in the weight vector. Since this can vary in the region R given in input, it is the same as considering a family of scoring function like $f$-skyline does. In light of this we can say that the concepts of $f$-dominance and $r$-dominance are equivalent. What UTK adds to $f$-skyline is the ability to also consider $k$ > 1. In the example 5.5 we considered k=1, and we see that the result coincides exactly with the tuples obtained by applying PO operator, i.e. p4 and p5, that are encountered by ray $w$ last (considering all the region R), and this is called 1-level of arrangement. However, if we consider k=2, UTK also returns the tuples that correspond to the lines that meet the ray $w$ in the region second last, so we have the 2-level of arrangement. Together they form the ≤ 2-level. In general, the ≤ $k$-level captures the top-$k$ set for any $w$ [13].

UTK the output depends on the (approximate) preferences specified by the user via a certain region R. Since in this approach the weight vector is expanded into a region, and report the additional options identified,it is providing providing the user with top-$k$ results for similar preference profiles. That's why all the requirements are achieved.

- **ORD-ORU:** to achieve personalization is used the linear scoring function that which is considered perfectly suitable in modeling human decision making [15]. In this approach is considered a preference vector $w$ which is incrementally expanded it equally in all directions in the preference domain. As the expansion radius grows, it gradually shifts towards standard dominance, including in the output additional records that cater to alternative preferences, similar to w. The stopping radius is indirectly (yet strictly) determined by the desired output size $m$. [12]. In this way ORD-ORU try to satisfy the three hard requirements.

Also in this case, as we did for UTK, we can notice that $f$-skyline is strictly related to this approach.

In ORD/ORU there is a preference vector, which is used in a scoring function which must be monotone like what $f$-skyline does. Since, we know that a record $p_i$ f-dominates a record $p_j$, if for all monotone functions belonging to the referred family of functions, $p_i$ scores at least as high as $p_j$, and strictly higher for at least one function. Thus, the concept of $\rho$-dominance, which states that a record $p_i$ $\rho$-dominates a record $p_j$ if, given a scoring function and preference vectors at $\rho$ distance from the seed vector, $p_i$ scores at least as high as $p_j$ for all preference vectors, and strictly higher for at least one of them, is equivalent to that of f-dominance.

If we focus on OSS, it is clear that the best choice is ORD-ORU, because in case of $f$-skyline or UTK, the user/application cannot determine the size of the output. On the contrary ORD-ORU returns an output of exactly $m$ records.

# 7 Conclusion

In this paper we analyzed three different approaches which aim is to solve the main problems in multiobjective settings: $f$-skyline, UTK and ORD-ORU.

At first we described the weakness of the standard skyline and top-$k$ queries, considering the notion of *dominance* and *ranking by utility*. Then an overview of the three approaches has been done, in order to explain, from a purely mathematical point of view, their behavior. The example, instead, try to make the explanations made clearer to the reader.

At the end, we have summarized the main aspects of the three approaches, to highlight how they seek to gap the bridge between personalization and OSS, and so between skyline and top-$k$ queries.